\documentclass[aps,prb,preprint]{revtex4}
\usepackage{graphicx}
\begin{document}



\newcommand*{\mycommand}[1]{\texttt{\emph{#1}}}

\author{Ahmad Mohammadi}
\affiliation{Laboratory of Physical Chemistry, ETH Zurich, 8093 Zurich, Switzerland}
\affiliation{Department of Physics, Persian Gulf University, 75196 Bushehr, Iran}
\author{Franziska Kaminski}
\altaffiliation{Niels Bohr Institute, 2100 Copenhagen, Denmark}
\author{Vahid Sandoghdar}
\author{Mario Agio}
\email{mario.agio@phys.chem.ethz.ch}
\affiliation{Laboratory of Physical Chemistry, ETH Zurich, 8093 Zurich, Switzerland}

\title{Fluorescence enhancement with the optical (bi-) conical antenna}

\begin{abstract}
We investigate the properties of finite gold nanocones as optical
antennas for enhancing molecular fluorescence. We compute the
modification of the excitation rate, spontaneous emission rate, and
quantum efficiency as a function of the nanocone base and length,
showing that the maximum field and fluorescence enhancements do not
occur for the same nanocone parameters. We compare the results with
those for nanorods and nanospheroids and find that nanocones perform
better.
\end{abstract}
\maketitle

\section{Introduction}

A series of experiments in the late 1960s and in the 1970s
discovered that molecular fluorescence and Raman scattering could be
enhanced close to metal
surfaces~\cite{drexhage66,drexhage68,drexhage74,jeanmaire77,albrecht77}.
These pioneering efforts spurred a fury of theoretical works in the
subsequent years to understand and explain the phenomena encountered
in surface-enhanced spectroscopy
\cite{chance78,gersten81,ruppin82,metiu84,moskovits85}. Among
various effects, these efforts identified field enhancement at sharp
edges and surface roughness as a source of stronger fluorescence
excitation and emission. It was also predicted that plasmon
resonances in metallic films and nanostructures could enhance these
effects further. Moreover, fluorescence quenching and the
modification of the emission quantum yield due to the absorption in
real metals were investigated theoretically. For all these effects,
one discovered a very strong influence on the emitter orientation
and position with respect to the metallic structure under study.
These steep dependencies, which are inherent to near-field
interactions, posed a great challenge to the experimental
verification and quantitative understanding of surface-enhanced
phenomena because they could not be controlled in ensemble
measurements. As a result, only the large enhancement factors had to
be deduced from averaging measurements.

The advent of scanning near-field optical microscopy
(SNOM)~\cite{pohl84,lewis84} and single molecule
spectroscopy~\cite{moerner89,orrit90} in the 1980s and their maturation in
the 1990s provided the experimental tools for controlled laboratory
investigation of surface-enhanced interactions. In 2004 we
succeeded, for the first time, to examine the enhancement of
fluorescence from a single oriented molecule as a function of its
precise three-dimensional (3D) location close to a single gold
nanoparticle. To do this, we placed a single gold nanosphere at the
end of a glass fiber tip~\cite{kalkbrenner01} and used scanning
probe microscopy to position it in the near field of single
molecules embedded in a thin film. In an extended series of
measurements we investigated the shortening of the fluorescence
lifetime, the modification of the emission spectrum, the change in
the emission pattern, dependence of the excitation rate on the
illumination wavelength within the plasmon resonance, and the size
of the gold
nanoparticle~\cite{kuehn06a,kuehn06b,kuehn08,eghlidi09}.
These studies showed that a spherical gold particle acts as a
nanoantenna that modifies the excitation rate, the spontaneous
emission rate, the emission spectrum, and the radiation pattern of
an emitter in its near field. Indeed, several other groups have also
investigated single molecule fluorescence in the near field of
well-defined scanning probes, which function as optical
antennas~\cite{frey2004,farahani05,anger06,taminiau08a}.

The concept of optical antennas is closely linked to the physics of
near-field microscopy, in which a tip mediates between a far-field
illumination and the sample in its subwavelength
vicinity~\cite{pohl00,greffet05}. Progress in nanofabrication has
motivated the realization of structures similar to radiowave
antennas,\cite{balanis05} for operating in the optical
domain~\cite{crozier03,muehlschlegel05,schuck05,li07,hofmann07,taminiau08a}
although various issues have to be considered. For example, metal
nanoparticles support localized surface plasmon-polariton (LSPP)
modes that depend on the nanoparticle shape, composition and
surroundings.\cite{bohren83} Moreover, losses due to absorption by
real metals are not negligible at optical
frequencies.\cite{CRCHandbook} In addition, the molecule is coupled
to the antenna via the displacement current, which is strongly
position and orientation dependent~\cite{ruppin82}. In this respect,
there are ongoing efforts to reconcile these differences with
standard antenna theory.\cite{novotny07,alu08}

Recently, we proposed a few simple rules for designing optical
antennas to enhance spontaneous emission by more than three orders
of magnitude while avoiding quenching\cite{rogobete07}. First, the
plasmon resonance should be in a spectral region where dissipation
in the metal is small. Second, the nanoparticle should have sharp
corners to strongly increase the near field. Third, the antenna and
the molecule dipole moment should be aligned in a head-to-tail
configuration to maximize coupling and radiation. Fourth, the
structure should be compatible with state-of-the-art
nanofabrication. We then performed a detailed analysis of nanorods
and nanospheroids for enhancing fluorescence by varying their aspect
ratio and composition. We found that nanospheroids perform better
than nanorods when the plasmon resonance needs to be shifted towards
shorter wavelengths\cite{mohammadi08b}. Moreover, by choosing
different plasmonic materials, such as aluminum, silver, copper and
gold, we could obtain large Purcell effects and quantum efficiencies
in a broad spectral region, from ultraviolet to near
infrared\cite{mohammadi09}. One of the issues that we identified is
that the antenna efficiency $\eta_\mathrm{a}$, which is the power
that reaches the far field divided by the total emitter power, and
the Purcell factor $F$, which represents the enhancement of the
radiative decay rate, are maximal for different antenna parameters.
In particular, we found that for gold nanorods and nanospheroids
increasing $\eta_\mathrm{a}$ involves a rapid decrease of $F$ in the
visible and near-infrared range\cite{mohammadi08b}.

The question now is if one can improve the antenna design to
increase $F$ without decreasing $\eta_\mathrm{a}$ and losing control
on the spectral position of the resonance. In this work, we propose
that a simple solution based on using a nanocone, where one end can
be sharp to increase the field enhancement and the Purcell factor,
while the other larger end increases the volume, and thus, the
antenna efficiency\cite{gersten80,gersten81}. The nanocone antenna
is similar to the (bi-) conical antenna, which is the canonical
example of a broadband antenna for applications in the VHF and UHF
frequency bands.\cite{balanis05} However, for practical reasons such
as the antenna weight, it is often realized in the form of a
bow-tie.

Finite and semi-infinite metal nanocones are not a new concept in
optics and
SNOM.\cite{betzig91,sanchez99,hartschuh03a,hartschuh03b,ichimura04}
For example, conical SNOM probes could be exploited to focus surface
plasmon-polaritons down to a spot size limited only by the tip
curvature.\cite{keilmann99,babadjanyan00,stockman04,vogel07}
Moreover, the field enhancement for
semi-infinite\cite{goncharenko06b} and
finite\cite{goncharenko06a,goncharenko07} silver nanocones has been
studied as a function of the cone angle. In both situations, like
for the bow-tie\cite{crozier03}, there exists an optimal angle that
maximizes the enhancement. Moreover, it was pointed out that a
finite nanocone can give rise to a stronger field than a
semi-infinite one because of the LSPP resonance
effect.\cite{martin01,krug02} Other groups investigated the
modification of fluorescence lifetime by semi-infinite metal
tips,\cite{kramer02,chang06,issa07a} where coupling to the surface
plasmon-polariton mode leads to quenching. Also for finite
nanocones, previous investigations concluded that quenching
dominates at the LSPP resonance.\cite{thomas04}

Here, we concentrate our attention on finite 3D gold nanocones
to demonstrate that they can actually exhibit very interesting performances
in terms of Purcell effect and antenna efficiency at the LSPP wavelength.
In particular, we discuss the role of $\eta_\mathrm{a}$
in determining the optimal angle for enhancing fluorescence.
First, we briefly review the theory and computational approach for
studying spontaneous emission and molecular fluorescence with an optical antenna.
Second, we investigate $F$ and $\eta_\mathrm{a}$
of single and double gold nanocones as a function of cone angle.
We then discuss the role of the emitter intrinsic quantum efficiency $\eta_o$,
which together with $F$ and $\eta_\mathrm{a}$ determines the effective quantum
efficiency and the fluorescence enhancement.
Lastly, we explore the effect of a supporting substrate and of rounding
the tip.

\section{Results and discussion}
\subsection{Theory and computational details}
\label{theory}

We consider an isolated emitter with radiative decay rate $\gamma_o^\mathrm{r}$,
nonradiative decay rate $\gamma_o^\mathrm{nr}$ and quantum efficiency
$\eta_o=\gamma_o^\mathrm{r}/(\gamma_o^\mathrm{r}+\gamma_o^\mathrm{nr})$.
The presence of a metal
nanostructure modifies the spontaneous emission rate and
introduces an additional non-radiative decay channel with rate $\gamma^\mathrm{nr}$
due to absorption in the metal.\cite{ruppin82,rogobete07}
The new radiative decay rate $\gamma^\mathrm{r}$
and the quantum efficiency $\eta$ can be related to the initial
values by the following expression\cite{mohammadi08b}
\begin{equation}
\eta=\frac{\eta_o}{(1-\eta_o)/F+\eta_o/\eta_\mathrm{a}},
\end{equation}
where $F=\gamma^\mathrm{r}/\gamma_o^\mathrm{r}$ is the Purcell factor
and $\eta_\mathrm{a}=\gamma^\mathrm{r}/(\gamma^\mathrm{r}+\gamma^\mathrm{nr})$ is the antenna efficiency.
Under weak excitation, the fluorescence signal of the isolated
emitter is $S_o=\xi_o\eta_o|\mathbf{d}\cdot{E}_o|^2$. Here, $\xi_o$
represents the collection efficiency, $\mathbf{d}$
is the transition electric dipole moment, and $\mathbf{E}_o$ is the electric field
at the emitter position. With an optical antenna,
the local electric field and the collection efficiency get also modified
and the signal becomes $S=\xi\eta|\mathbf{d}\cdot\mathbf{E}|^2$.\cite{kuehn06a,kuehn08}
Assuming that the signal is collected over all angles,
such that $\xi=\xi_o=1$, the fluorescence
enhancement reads
\begin{equation}
\label{fluo1}
\frac{S}{S_o}=\frac{\eta}{\eta_o}
\frac{|\mathbf{d}\cdot\mathbf{E}|^2}{|\mathbf{d}\cdot\mathbf{E}_o|^2}.
\end{equation}
Furthermore, if the metal nanostructure and the emitter are arranged
in a configuration that almost preserves the dipolar radiation pattern
of the isolated emitter,\cite{kuehn08,mohammadi08b}
reciprocity implies that $S/S_o$
can be well approximated by replacing the electric
field enhancement $|\mathbf{d}\cdot\mathbf{E}|^2/|\mathbf{d}\cdot\mathbf{E}_o|^2$
with the Purcell factor $F$~\cite{taminiau08c}. We thus write
\begin{equation}
\label{fluo2}
\frac{S}{S_o}\simeq\frac{F}{(1-\eta_o)/F+\eta_o/\eta_\mathrm{a}}.
\end{equation}
From Eqs.~(\ref{fluo1}) and (\ref{fluo2}) one immediately notes that the maximum
fluorescence and the maximum field enhancements do not correspond because
of $\eta_o$ and $\eta_\mathrm{a}$.

The calculation of $F$ and $\eta_\mathrm{a}$ is performed here using the
body-of-revolution (BOR) finite-difference time-domain (FDTD)
method, which exploits the cylindrical symmetry of the nanocones
to significantly reduce the computational
burden in comparison to standard 3D-FDTD methods\cite{taflove05}. The computational details can be found
in Ref.~\onlinecite{mohammadi08b} and references therein.
We focus our attention on gold\cite{CRCHandbook} optical antennas made
of one or two finite nanocones. The emitter is always at 10 nm from the
sharp end of the nanocone, unless otherwise specified,
and it is positioned and oriented along the nanocone axis.
This distance is chosen in order to ignore effects due
to nonlocality in the optical constants of the metal interface\cite{ford84}
and convergence issues in the FDTD method.\cite{kaminski07}
The mesh pitch is 1 nm or 0.5 nm, depending on the nanocone geometry.
For the case of an optical antenna made of two nanocones, the gap
between them is fixed to 20 nm.
The nanocone dimensions are chosen such that the LSPP is
placed in the visible and near-infrared spectral range.
The surrounding material is either glass (refractive index 1.5) or air.

\subsection{Single and double conical optical antennas}
\label{sdcone}

\begin{figure}
\includegraphics[width=16cm]{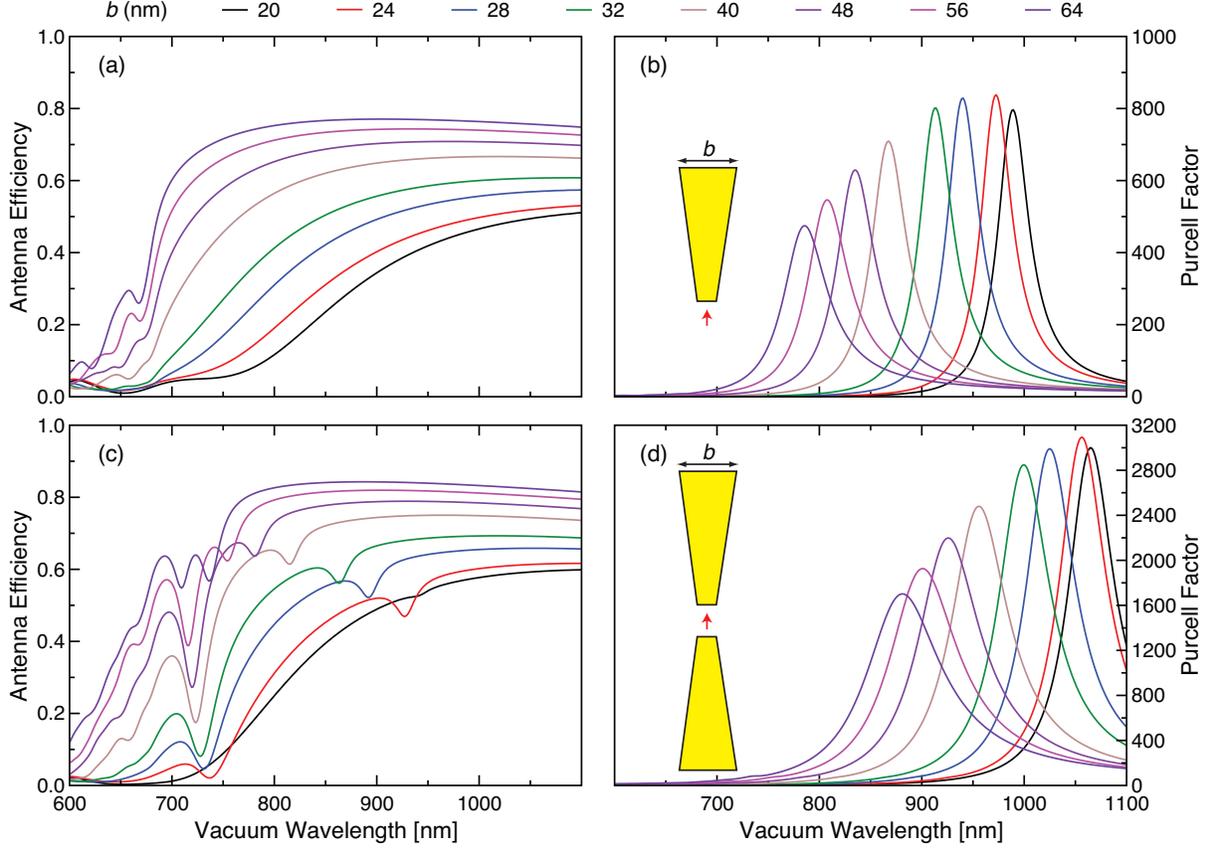}
\caption{\label{smallcone}Single, (a) and (b), and double,
(c) and (d), conical optical antennas in glass
(refractive index 1.5). Purcell factor $F$, (b) and (d), and antenna
efficiency $\eta_\mathrm{a}$, (a) and (c), as a function of the base diameter $b$.
The nanocone is 80 nm long and it has a fixed tip diameter of 20 nm.
The optical antenna gap for the case of two nanocones is 20 nm.
The insets in (b) and (d) sketch the antenna cross section, indicating the
emitter position and orientation, and the nanocone parameter that is varied.}
\end{figure}

\ref{smallcone} displays the Purcell factor $F$
and the antenna efficiency $\eta_\mathrm{a}$ for single and double nanocone antennas
in glass as a function of the nanocone base $b$ and the wavelength.
The nanocone height is 80 nm and the tip end is flat with a fixed
diameter of 20 nm.
The black curves correspond to the case of a nanorod\cite{mohammadi08b}.
Starting from this structure, the Purcell factor increases slightly and then
decreases, confirming that there exists an optimal value for
$b$\cite{goncharenko06a,goncharenko07}.
This trend is found for optical antennas made of one and two nanocones,
as shown in \ref{smallcone}(b) and \ref{smallcone}(d).
However, \ref{smallcone}(a) and \ref{smallcone}(c) reveal that $\eta_\mathrm{a}$
steadily grows with $b$ because the volume of the nanostructure
increases.
Intuitively, that occurs since scattering is proportional
to the volume squared, while absorption only to the volume\cite{bohren83}.
Therefore, if both F and $\eta_\mathrm{a}$ have to be
considered in the assessement of the antenna performances,
one sees that $b$ might be different than the optimal
value found when only the field enhancement is taken
into account.
The dips in \ref{smallcone}(a) and \ref{smallcone}(c) are associated
with the excitation of higher-order resonances, which increase
the non-radiative decay rate more strongly than
the radiative one~\cite{rogobete07}. 
An important advantage with respect to nanorods and nanospheroids is
that here the resonance can be tuned towards the visible spectrum
simply by changing the nanocone angle, without a significant
loss of enhancement. Indeed, for nanorods we found
that shifting the resonance by 100 nm can lead to a reduction
of $F$ by more than a factor of two.\cite{mohammadi08b}

\begin{figure}
\includegraphics[width=16cm]{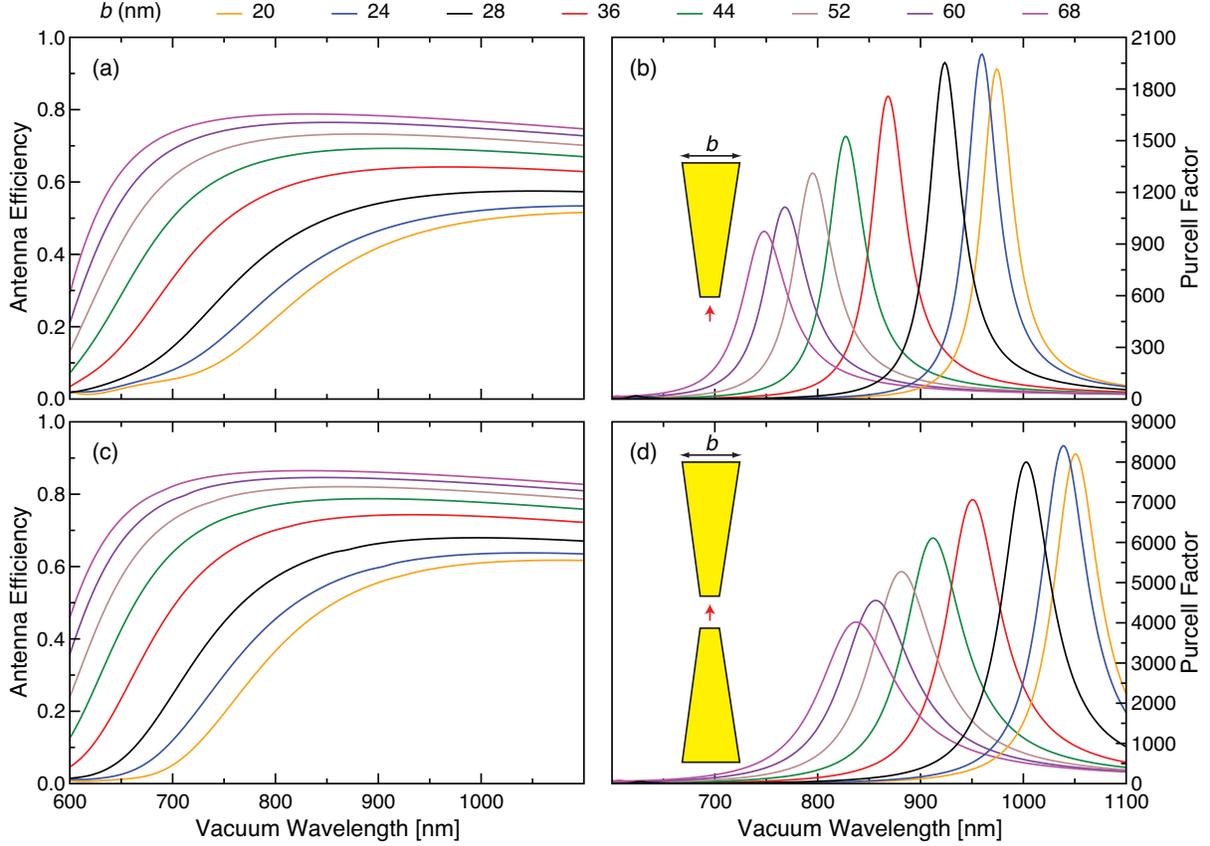}
\caption{\label{largecone}Single, (a) and (b), and double,
(c) and (d), conical optical antennas in air
(refractive index 1.0). Purcell factor $F$, (b) and (d), and antenna
efficiency $\eta_\mathrm{a}$, (a) and (c), as a function of the base diameter $b$.
The nanocone is 140 nm long and it has a fixed tip diameter of 20 nm.
The optical antenna gap for the case of two nanocones is 20 nm.
The insets in (b) and (d) sketch the antenna cross section, indicating the
emitter position and orientation, and the nanocone parameter that is varied.}
\end{figure}

In \ref{largecone} we consider another set of optical antennas, where the surrounding medium
is air and the nanocone height is 140 nm. The redshift due to a longer
optical antenna is compensated by a lower refractive index so that
the resonances are in the same spectral region of those shown
in \ref{smallcone}. The overall behavior as a function of $b$
is similar to the previous case.  The main difference here
is that the Purcell factor rises to as high as 2000 for a single
nanocone and 8000 for a double nanocone antenna.
Such large values combined with a very good antenna efficiency
and a wide spectral tunability of the LSPP make them ideal
systems for enhancing the radiative properties of solid-state quantum emitters.

\subsection{Role of the initial quantum efficiency}
\label{etao}

\begin{figure}
\includegraphics[width=8.25cm]{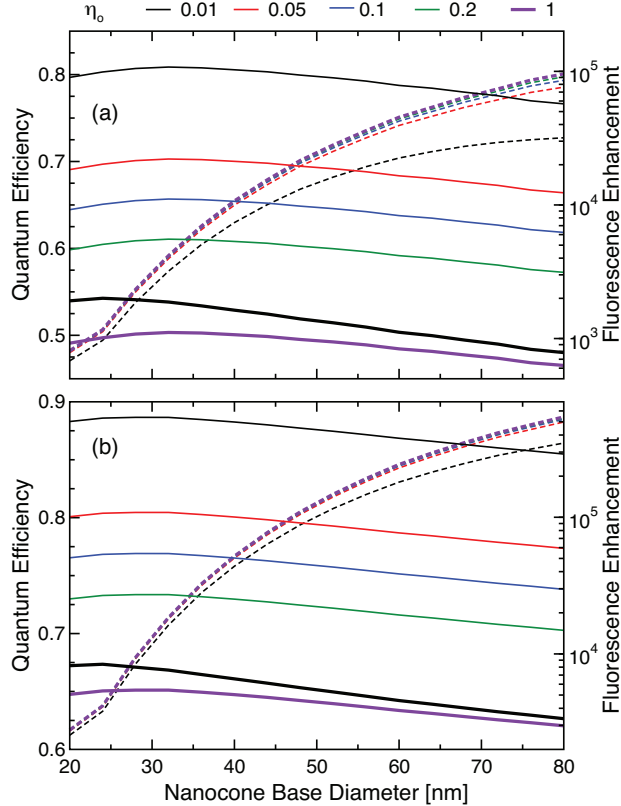}
\caption{\label{fluocone}Single (a) and double (b) conical optical antennas in air
(refractive index 1.0). Fluorescence enhancement $S/S_o$ (solid curves)
and quantum efficiency $\eta$ (dashed curves) as a function of the base diameter $b$
at the wavelength corresponding to the maximal Purcell factor.
The bold black solid curve represents the Purcell factor.
The nanocone is 140 nm long and it has a fixed tip diameter of 20 nm.
The optical antenna gap for the case of two nanocones is 20 nm.
See the inset to \ref{largecone} for details on the coupling geometry.}
\end{figure}

According to Eq.~(\ref{fluo2}), the fluorescence enhancement
depends on $F$, on $\eta_\mathrm{a}$, and on the initial quantum
efficiency $\eta_o$. We anticipated that the value of
$b$ that maximizes $F$ should not correspond to the value that
maximizes $S/S_o$. To discuss this point further we plot in \ref{fluocone}
the fluorescence enhancement as a function of the nanocone base
diameter $b$ and $\eta_o$. For each $b$ we
choose the wavelength associated with the maximum $F$.
For the same wavelength, we also take $\eta_\mathrm{a}$ and use these quantities
in Eq.~(\ref{fluo2}).  For a given $\eta_o$, $S/S_o$ is maximal when
$b$ is about 30 nm in place of 24 nm obtained for $F$ (see the bold
black curve). The difference is small because $F$ is so large that
Eq.~(\ref{fluo2}) can be well approximated by $S/S_o=F\eta_\mathrm{a}/\eta_o$.
Since $\eta_\mathrm{a}$ does not change much compared to $F$,
it turns out that maximizing $F$ and $S/S_o$ give similar
values for $b$. This can also be inferred from the quantum
efficiency $\eta$, which is shown by dashed curves in \ref{fluocone}.
$\eta$ is close to $\eta_\mathrm{a}$, corresponding to the curve
for $\eta_o=100$\%, even when $\eta_o=5$\%.
For smaller Purcell factors, the difference
between the optimal $b$ for $F$ and for $S/S_o$ would
be larger and vary with $\eta_o$.

\subsection{Effect of a supporting substrate}
\label{substrate}

\begin{figure} \includegraphics[width=8.25cm]{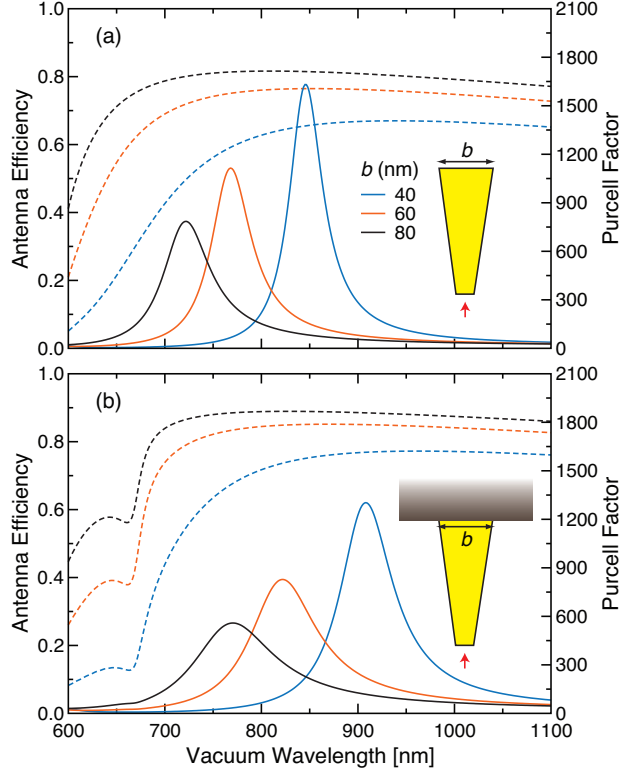}
\caption{\label{subs}Effect of a supporting substrate on nanocone
optical antennas in air. Purcell factor $F$ (solid curves) and antenna
efficiency $\eta_\mathrm{a}$ (dashed curves) without (a) and with (b) a glass substrate
(refractive index 1.5). The nanocone is 140 nm long, it has a base
diameter of 40, 60 or 80 nm and a fixed tip diameter of 20 nm.
The insets in (a) and (b) sketch the antenna cross section, indicating the
emitter position and orientation, and the nanocone parameter that is varied.}
\end{figure}

An important aspect for the experimental realization of nanoantennas
is the effect of a supporting substrate. Often, the optical antenna
is grown on a dielectric substrate,\cite{stade07,fredriksson07,kim08,fleischer09}
or is attached to the end of a fiber.\cite{kuehn06a,fleischer08,deangelis08,zou09}
Previous works concluded that the presence of a substrate has
a negligible effect on the field enhancement and on the spectral
position of the LSPP for nanocones.\cite{goncharenko06a,goncharenko07}
Here, in \ref{subs}, we show that adding a glass substrate can shift the
resonance by more than 50 nm. Furthermore, the LSPP exhibits a
stronger radiative broadening, which decreases the Purcell factor $F$ and
the field enhancement,\cite{meier83} and increases the
antenna efficiency $\eta_\mathrm{a}$. Note that the shift
is smaller for the nanocone having the largest base diameter.

\subsection{Effect of the tip termination}
\label{tipsec}

\begin{figure}
\includegraphics[width=8.25cm]{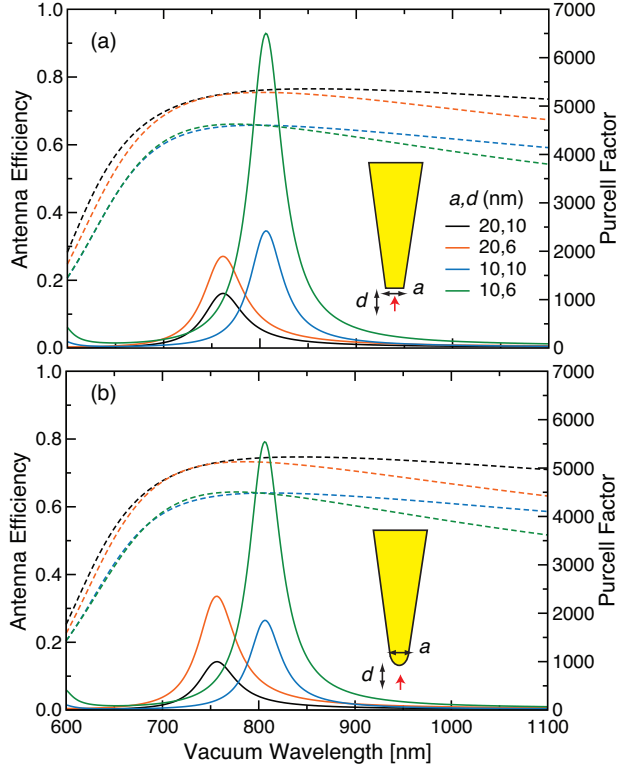}
\caption{\label{tip}Effect of the tip termination on nanocone optical antennas
in air. Purcell factor $F$ (solid curves) and antenna efficiency $\eta_\mathrm{a}$
(dashed curves) for a flat (a) and rounded (b) tip.
The nanocone is 140 nm long, it has a fixed base diameter of 60 nm
and a tip diameter of 10 or 20 nm. The emitter is at a distance
of 6 or 10 nm from the end of the tip.
The insets in (a) and (b) sketch the antenna cross section, indicating the
emitter position and orientation, and the varied nanocone parameters.}
\end{figure}

The last important aspect that we would like to address
is the effect of the tip termination.
Previous studies on the field enhancement in triangular nanoparticles
have shown that changing the tip termination can affect
the results.\cite{kottmann01,hao04}
Here, we compare the case of a flat tip, shown in \ref{tip}(a),
with the case of a rounded tip having the same diameter,
as shown in \ref{tip}(b).
We see that the resonance frequency and the antenna efficiency
$\eta_\mathrm{a}$ are almost the same. The only noticeable difference occurs
for the Purcell factor $F$, which is larger for the case of a flat tip.
This holds also if the flat and rounded ends have a diameter
of 10 nm in place of 20 nm.
Nevertheless, the difference between the two situations
is not huge. Indeed, when the emitter is only 6 nm from the
metal surface, the 20 nm rounded tip yields a stronger field enhancement
in comparison to the flat one. These results stem essentially from the
complex behavior of the electric field close to a metal tip~\cite{novotny95b}.

\section{Conclusion}

We investigated the performance of gold nanocones as optical
antennas for enhancing molecular fluorescence. Compared to nanorods
and nanospheroids, the great advantage of nanocones is that the
spectral position of the LSPP can be tuned towards shorter
wavelengths by increasing the cone angle without compromising the
Purcell factor and reducing the antenna efficiency. Another
important practical aspect in favor of nanocones is the increased
robustness and stability for vertical orientation because the base
is larger than for
nanorods.\cite{stade07,fredriksson07,kim08,fleischer09,fleischer08,deangelis08,zou09}

Moreover, we have shown that an optimal
angle for the Purcell factor exists, which is directly
related to the field enhancement studied
in previous works.\cite{goncharenko06b,goncharenko06a,goncharenko07}
We have also studied the antenna efficiency, showing that quenching
does not necessarily occur in correspondence with the LSPP.\cite{thomas04}
While $F$ increases and then decreases with increasing cone angle,
$\eta_\mathrm{a}$ exhibits a monotonic growths. Therefore,
because the fluorescence signal depends on the local electric field and on
the quantum efficiency, the maximum field and fluorescence enhancements
do not occur for the same cone angle.
Nevertheless, we found that this difference is small
if the Purcell factor is much larger than 1, which is easy to
achieve using realistic nanocone parameters.

The strong fluorescence and field enhancement as well as the
electric field localization at the nanocone tip holds great promise
for high-resolution fluorescence, Raman, and other nonlinear
nanoscopies.\cite{sanchez99,hartschuh03a,hartschuh03b,ichimura04} In
fact, finite nanocones could be very efficiently butt-coupled to a
dielectric nanofiber or interfaced with a weakly-focused radially-polarized
beam for the realization of a high-throughput and
large-bandwidth scanning near-field optical microscope\cite{chen09,chen10}.
Considering the recent progress in the fabrication of single and
double gold nanocone on
substrates,\cite{stade07,fredriksson07,kim08,fleischer09}
cantilevers,\cite{deangelis08,zou09} and on the facet of optical
fibers,\cite{fleischer08} we anticipate a great deal of
activity on this antenna system.

The authors would like to thank X.-W. Chen, E. Di Fabrizio, F. De Angelis,
H. M. Eghlidi, K.-G. Lee and S. G\"otzinger for fruitful discussions.
A. Mohammadi acknowledges support from
the Persian Gulf University Research Council. This work was performed
in the framework of the ETH Zurich initiative on Composite Doped
Metamaterials (CDM).



\end{document}